\documentclass[aps,prd,superscriptaddress,twocolumn,showpacs]{revtex4}
\usepackage{graphicx}
\usepackage{epstopdf}
\usepackage{amsmath}
\usepackage{amsfonts}
\usepackage{amssymb}
\usepackage{latexsym}
\usepackage{verbatim}

\begin{document}

\title{Remarks on an anomalous triple gauge boson couplings}
\author{Patricio Gaete}
\email{patricio.gaete@usm.cl}
\affiliation{Departmento de F\'{\i}sica and Centro Cient\'{\i}fico-Tecnol\'ogico de
Valpara\'{\i}so-CCTVal, Universidad T\'{e}cnica Federico Santa Mar\'{\i}a, Valpara\'{\i}so, Chile}
\author{J. A. Helay\"el-Neto}
\email{helayel@cbpf.br}
\affiliation{Centro Brasileiro de Pesquisas F\'\i sicas, Rio de Janeiro, RJ, Brasil}
\author{L. P. R. Ospedal}
\email{ leoopr@cbpf.br}
\affiliation{Centro Brasileiro de Pesquisas F\'\i sicas, Rio de Janeiro, RJ, Brasil}
\date{\today }

\begin{abstract}
We address the effect of an anomalous triple gauge boson couplings on a physical observable for the electroweak sector of the Standard Model, when the $SU(2)_{L}\otimes U(1)_{Y}$ symmetry is spontaneously broken by the Higgs mechanism to $U(1)_{em}$. Our calculation is done within the framework of the gauge-invariant, but path-dependent variable formalism which is alternative to the Wilson loop approach. Our result shows that the interaction energy is the sum of a Yukawa and a linear potential, leading to the confinement of static probe charges. The point we wish to emphasize, however, is that the anomalous triple gauge boson couplings ($Z \gamma \gamma$) contributes to the confinement for distances on the intranuclear scale.
\end{abstract}

\maketitle

\section{Introduction}

It is well known that the Standard Model (SM) is an active field of research, based on the gauge group $SU(3)_{C}\otimes SU(2)_{L}\otimes U(1)_{Y}$, which has been successful in describing many of the particle physics phenomena. However, despite this great success, the SM must be extended to explain some aspects that need to be understood. Probably the most striking examples are: dark matter candidates, nonzero neutrino masses, baryon asymmetry of the Universe and origin of the electroweak scale. These puzzling facts have led to an increasing interest in physics beyond the SM (BSM). Mention should be made, at this point, to the Higgs boson discovered at the LHC \cite{Atlas,CMS} which clearly corroborated the electroweak symmetry breaking. In other words, the $SU(2)_{L}\otimes U(1)_{Y}$ symmetry is spontaneously broken by the Higgs mechanism \cite{Englert,Higgs1,Higgs2} to $U(1)_{em}$. However a full understanding of this mechanism from first principles still remain elusive.

It is also known that considerable attention has been paid to the investigation of anomalous triple gauge boson couplings of the electroweak sector \cite{Rahaman:2018ujg,Senol:2018cks,Senol:2019qyl,Larios:2000ni,Degrande:2013kka,Hagiwara:1986vm,Gounaris:1999kf,Green:2016trm}. The interest in studying these couplings is mainly due to the possibility of providing a better understanding of the electroweak symmetry breaking mechanism and test the predictions in collider experiments. It is believed that the presence of these couplings  can give important hints of new physics beyond the SM. In this connection, it may be recalled that the $ZZ$ and $Z\gamma$ production are the foremost processes where the triple couplings between the usual photon and $Z$ boson ($Z \gamma \gamma$ and $Z \gamma Z$) can be studied. This can be achieved by adding higher dimension effective operators to the Lagrangian of the SM. For example, in this work we will consider an interaction term of the form \cite{Rahaman:2018ujg}
\begin{equation}
{{\cal L}_{Int}} =  - \frac{e}{{m_Z^2}}h_3^\gamma \left( {{\partial _\sigma }{F^{\sigma \rho }}} \right){Z^\alpha }{\tilde F_{\rho \alpha }}, \label{wc-01}
\end{equation}
where $h_3^\gamma$ appears in the $Z\gamma$ production process.

Motivated by these observations and given experimental data on $ZZ$ and $Z\gamma$ production, it is desirable to have some additional understanding of the physical consequences presented by these anomalous triple gauge boson couplings of the electroweak sector. Of particular concern to us is the effect of the interaction term  on a physical observable. To do this, we will work out the static potential for the theory under consideration by using the gauge-invariant but path-dependent variables formalism \cite{Gaete:2017vxk,Gaete:2017cpc}. According this formalism, the interaction energy between two static charges is obtained once a judicious identification of the physical degrees of freedom is made. It also provides an alternative technique for determining the static potential for a gauge theory. Interestingly enough, the static potential profile contains a linear term, leading to the confinement of static probe charges. 

Our work is organized according to the following outline: in Sect. 2, we analyze the interaction energy and force for a fermion-antifermion pair. In Sect. 3, we discuss our results. Finally, in Sect. 4, we cast our  Final Remarks. 

In our conventions the signature of the metric is $(+1,-1,-1,-1)$.

\section{Interaction energy and Force}

As already expressed, the gauge theory we are considering describes the interaction between the familiar massless $U(1)_{em}$ photon with the massive vector $Z$-field via a new coupling. In this case, the corresponding Lagrangian density takes the form:
\begin{equation}
{\cal L} =  - \frac{1}{4}F_{\mu \nu }^2 - \frac{1}{4}Z_{\mu \nu }^2 + \frac{1}{2}m_Z^2{Z_\mu }{Z^\mu } + \chi  \left( {{\partial ^\alpha }{F_{\alpha \mu }}} \right){\tilde F^{\mu \beta }}{Z_\beta }, 
\label{wc-05}
\end{equation}
where $m_{Z}$ is the mass for the gauge boson $Z$, ${\tilde F_{\mu \nu }} = {\textstyle{1 \over 2}}{\varepsilon _{\mu \nu \lambda \rho }}{F^{\lambda \rho }}$, and $\chi =  - \frac{e}{{m_Z^2}}h_3^\gamma$ represents the coupling constant. 

It is of interest also to notice that if we consider the foregoing model in the limit of a very heavy $Z$-field (and we are bound to energies much below $m_{Z}$) we are allowed to integrate over $Z_{\mu}$, which, then, yields an effective theory for the $A_{\mu}$-field. This can be readily accomplished by means of the path integral formulation of the generating functional associated to Eq.(\ref{wc-05}).
Once this is done, we find that the effective theory can be brought to the form:
\begin{eqnarray}
{\cal L} &=&  - \frac{1}{4}F_{\mu \nu }^2 - \frac{{{\chi ^2}}}{2}\left( {{\partial ^\alpha }{F_{\alpha \beta }}} \right){\tilde F^{\beta \mu }}\frac{1}{{\left( {\Delta  + m_Z^2} \right)}}\left( {{\partial _\gamma }{F^{\gamma \delta }}} \right){\tilde F_{\delta \mu }} \nonumber\\
&-& \frac{{{\chi ^2}}}{2}\left( {{\partial ^\alpha }{F_{\alpha \beta }}} \right){\tilde F^{\beta \mu }}\frac{{{\partial _\mu }{\partial _\nu }}}{{m_Z^2\left( {\Delta  + m_Z^2} \right)}}\left( {{\partial ^\gamma }{F_{\gamma \delta }}} \right){\tilde F^{\delta \nu }},                \label{wc-10}
\end{eqnarray}
where $\Delta  \equiv {\partial _\mu }{\partial ^\mu }$.
It may be remarked in passing that the above Lagrangian density is a theory with non-local time derivatives.
However, we stress that this paper is aimed at studying the static potential, so that $\Delta$ can be replaced by $-{\nabla ^2}$. For notational convenience we have maintained $\Delta$, but it should be borne in mind that this paper essentially deals with the static case. 

We next observe that, in order to study quantum properties of the electromagnetic field in the presence of external electric and magnetic fields, we should split the ${A_\mu }$-field as the sum of a classical background, $\left\langle {{A_\mu }} \right\rangle$, and a small quantum fluctuation, ${a_\mu }$, namely: ${A_\mu } = \left\langle {{A_\mu }} \right\rangle  + {a_\mu }$. Making use of this expression, we find that Eq. (\ref{wc-10}), up to quadratic terms in the fluctuations, reduces to
\begin{eqnarray}
{\cal L} &=&  - \frac{1}{4}{f_{\mu \nu }}\left[ {1 + \frac{{{\chi ^2}}}{2}\left\langle {{F_{\alpha \beta }}} \right\rangle \left\langle {{F^{\alpha \beta }}} \right\rangle \frac{\Delta }{{\left( {\Delta  + m_Z^2} \right)}}} \right]{f^{\mu \nu }} \nonumber\\
&-& \frac{{{\chi ^2}}}{2}\left\langle {{F_{\delta \tau }}} \right\rangle \left\langle {{F^{\tau \beta }}} \right\rangle {f_{\alpha \beta }}\frac{{{\partial ^\alpha }{\partial _\beta }}}{{\left( {\Delta  + m_Z^2} \right)}}{f^{\gamma \delta }}, \label{wc-15} 
\end{eqnarray}
where ${f_{\mu \nu }} = {\partial _\mu }{a_\nu } - {\partial _\nu }{a_\mu }$.
It should, however, be noted here that by defining ${v_{\alpha \beta }} = \frac{1}{2}{\varepsilon _{\alpha \beta \mu \nu }}\left\langle {{F^{\mu \nu }}} \right\rangle$, the foregoing equation becomes
\begin{eqnarray}
{\cal L} &=&  - \frac{1}{4}{f_{\mu \nu }}\left[ {1 - {\chi ^2}{v^{\rho \lambda }}{v_{\rho \lambda }}\frac{\Delta }{{\left( {\Delta  + m_Z^2} \right)}}} \right]{f^{\mu \nu }} \nonumber\\
&-& \frac{{{\chi ^2}}}{2}{v^{\lambda \beta }}\left( {{\partial ^\alpha }{f_{\alpha \beta }}} \right)\frac{1}{{\left( {\Delta  + m_Z^2} \right)}}{v_{\delta \lambda }}\left( {{\partial _\gamma }{f^{\gamma \delta }}} \right). \label{wc-20} 
\end{eqnarray}
And, finally, by considering the ${v^{0i}} \ne 0$ and ${v^{ij}} = 0$ case (referred to as the magnetic one in what follows), we readily deduce that 
\begin{eqnarray}
{\cal L} &=&  - \frac{1}{4}{f_{\mu \nu }}\left[ {1 + 2{\chi ^2}{{\bf v}^2}\frac{{{\nabla ^2}}}{{\left( {{\nabla ^2} - m_Z^2} \right)}}} \right]{f^{\mu \nu }} \nonumber\\
&-& \frac{{{\chi ^2}}}{2}{{\bf v}^2}{f_{i0}}\frac{{{\partial ^i}{\partial _j}}}{{\left( {{\nabla ^2} - m_Z^2} \right)}}{f^{j0}} \nonumber\\
&+& \frac{{{\chi ^2}}}{2}{v^{j0}}{v_{k0}}{f_{ij}}\frac{{{\partial ^i}{\partial _l}}}{{\left( {{\nabla ^2} - m_Z^2} \right)}}{f^{lk}}. \label{wc-25}
\end{eqnarray}

We are now in a position to evaluate the corresponding interaction energy in the case under consideration. To do that we now carry out a Hamiltonian analysis of this theory.

Let us start by observing that the canonical Hamiltonian is given by
\begin{eqnarray}
{H_C} &=& \int {{d^3}x} \left\{ {{\Pi _i}{\partial ^i}{A_0} - \frac{1}{2}{\Pi ^i}\frac{{\left( {{\nabla ^2} - m_Z^2} \right)}}{{\left( {\alpha {\nabla ^2} - m_Z^2} \right)}}{\Pi ^i}} \right\} \nonumber\\
&-& \frac{1}{2}\int {{d^3}x} \left\{ {{\Pi ^i}\frac{{\left( {{\nabla ^2} - m_Z^2} \right)}}{{\left( {{\nabla ^2} - {\Omega ^2}} \right)\left( {\alpha {\nabla ^2} - m_Z^2} \right)}}{\partial _i}{\partial _k}{\Pi ^k}} \right\}  \nonumber\\
&+& 2\int {{d^3}x} \left\{ {{\Pi ^i}\frac{{\left( {{\nabla ^2} - m_Z^2} \right)}}{{{\Omega ^2}\left( {{\nabla ^2} - {\Omega ^2}} \right)\left( {\alpha {\nabla ^2} - m_Z^2} \right)}}{\partial _i}{\partial _k}{\Pi ^k}} \right\} \nonumber\\
&+& \frac{1}{4}\int {{d^3}x} \left\{ {{f_{ij}}\frac{{\left( {\alpha {\nabla ^2} - m_Z^2} \right)}}{{\left( {{\nabla ^2} - m_Z^2} \right)}}{f^{ij}}} \right\} \nonumber\\
&-& \frac{{{\chi ^2}}}{2}\int {{d^3}x} \left\{ {{v^{j0}}{v_{ko}}{f_{ij}}\frac{{{\partial ^i}{\partial _l}}}{{\left( {{\nabla ^2} - m_Z^2} \right)}}{f^{lk}}} \right\}, \label{wc-30}
 \end{eqnarray}
where $\alpha  = 1 + 2{\chi ^2}{{\bf v}^2}$, whereas $\frac{1}{{{\Omega ^2}}} = \frac{{{\chi ^2}{{\bf v}^2}}}{{\left( {\alpha {\nabla ^2} - m_Z^2} \right)}}$.

Time conservation of the primary constraint, $\Pi_0$, yields a secondary constraint. The secondary constraint is therefore the usual Gauss's law, ${\Gamma _1} \equiv {\partial _i}{\Pi ^i} = 0$, and together displays the first-class structure of the theory. The extended (first-class) Hamiltonian that generates the time evolution of the dynamical variables then reads  $H = H_C + \int {d^3 } x\left( {c_0 \left( x \right)\Pi _0 \left( x \right) + c_1
\left( x\right)\Gamma _1 \left( x \right)} \right)$, where $c_0 \left( x
\right)$ and $c_1 \left( x \right)$ are arbitrary functions of space and time.
It may be noted here that $\Pi^0 = 0$ for all time 
and $\dot{A}_0 \left( x \right)= \left[ {A_0\left( x \right),H} \right] = c_0 \left( x \right)$, which
is completely arbitrary. We may accordingly discard $A^0 $ and $\Pi^0$. It is of interest also to notice that it is redundant to retain the term containing $A_0$ because it can be absorbed by redefining the function 
$c_1 (x)$. From the above, the extended Hamiltonian is then
\begin{eqnarray}
H &=& \int {{d^3}x} \left\{ {c\left( x \right){\partial _i}{\Pi ^i} - \frac{1}{2}{\Pi ^i}\frac{{\left( {{\nabla ^2} - m_Z^2} \right)}}{{\left( {\alpha {\nabla ^2} - m_Z^2} \right)}}{\Pi ^i}} \right\} \nonumber\\
 &+& \frac{{{\chi ^2}{{\bf v}^2}}}{2}\int {{d^3}x} \left\{ {{\Pi ^i}\Theta {\partial _i}{\partial _k}{\Pi ^k}} \right\} \nonumber\\ 
&-& 2{\chi ^2}{{\bf v}^4}\int {{d^3}x} \left\{ {{\Pi ^i}\Xi {\partial _i}{\partial _k}{\Pi ^k}} \right\}, 
\label{wc-35}
\end{eqnarray}
where
\begin{equation}
\Theta  = \frac{{\left( {{\nabla ^2} - m_Z^2} \right)}}{{\left( {\alpha  - {\chi ^2}{{\bf v}^2}} \right)\left( {{\nabla ^2} - m_Z^2} \right)\left( {\alpha {\nabla ^2} - m_Z^2} \right)}}, \label{wc-35a}
\end{equation} 
\begin{equation}
\Xi  = \frac{{\left( {{\nabla ^2} - m_Z^2} \right)}}{{\left[ {\left( {\alpha  - {\chi ^2}{{\bf v}^2}} \right)\left( {{\nabla ^2} - m_Z^2} \right)} \right]{{\left( {\alpha {\nabla ^2} - m_Z^2} \right)}^4}}}, \label{wc-35b}
\end{equation}
and $c(x) = c_1 (x) - A_0 (x)$. 

We can at this stage impose a gauge fixing condition that together with the first class constraint, ${\Gamma _1}(x)$, the full set of constraints become second class. We consequently choose the gauge fixing condition as \cite{Gaete:1997eg}:
 \begin{equation}
\Gamma _2 \left( x \right) \equiv \int\limits_{C_{\zeta x} } {dz^\nu }
A_\nu\left( z \right) \equiv \int\limits_0^1 {d\lambda x^i } A_i \left( {
\lambda x } \right) = 0,  \label{wc-40}
\end{equation}
where $\lambda$ $(0\leq \lambda\leq1)$ is the parameter describing the
space-like straight path $x^i = \zeta ^i + \lambda \left( {x - \zeta}
\right)^i $ , and $\zeta $ is a fixed point (reference point). In passing we observe
that there is no essential loss of generality if we restrict our considerations to $\zeta^i=0 
$. Hence, we readly find that the only non-vanishing equal-time Dirac bracket reads
\begin{eqnarray}
\left\{ {A_i \left( {\bf x} \right),\Pi ^j \left( {\bf y} \right)} \right\}^ * &=&\delta{\
_i^j} \delta ^{\left( 3 \right)} \left( {{\bf x} - {\bf y}} \right) \nonumber\\
&-& \partial _i^x 
\int\limits_0^1 {d\lambda x^j } \delta ^{\left( 3 \right)} \left( {\lambda
{\bf x}- {\bf y}} \right).  \label{wc-45}
\end{eqnarray}

We now pass on to the calculation of the interaction energy. To do this we compute the expectation value of the energy operator H in the physical state $\left| \Phi  \right\rangle$. We next observe that the physical state $\left| \Phi  \right\rangle$ can be written as 
\begin{eqnarray}
\left| \Phi  \right\rangle  &\equiv& \left| {\bar \Psi \left( {\bf y} \right)\Psi \left( {{{\bf y}^ {\prime} }} \right)} \right\rangle  \nonumber\\ 
&=& \bar \Psi \left( {\bf y} \right)\exp \left( {iq\int_{{{\bf y}^ {\prime} }}^{\bf y} {d{z^i}{A_i}\left( z \right)} } \right)\Psi \left( {{{\bf y}^ {\prime} }} \right)\left| 0 \right\rangle,   \label{wc-50}
\end{eqnarray}
where the line integral is along a spacelike path on a fixed time slice, $q$ is the fermionic charge and $\left| 0 \right\rangle$ is the physical vacuum state.

It is also important to observe that taking the above Hamiltonian structure into account, one encounters
\begin{eqnarray}
{\Pi _i}\left( {\bf x} \right)\left| {\bar \Psi \left( {\bf y} \right)\Psi \left( {{{\bf y}^ {\prime} }} \right)} \right\rangle  &=& \bar \Psi \left( {\bf y} \right)\Psi \left( {{{\bf y}^ {\prime} }} \right){\Pi _i}\left( {\bf x} \right)\left| 0 \right\rangle \nonumber\\ 
&+& q\int_{\bf y}^{{{\bf y}^ {\prime} }} {d{z_i}{\delta ^{\left( 3 \right)}}\left( {{\bf z} - {\bf x}} \right)\left| \Phi  \right\rangle }. \nonumber\\
\label{wc-55}
\end{eqnarray}
Thus, we obtain for ${\left\langle H \right\rangle _\Phi }$ the expression 
\begin{equation}
{\left\langle H \right\rangle _\Phi } = {\left\langle H \right\rangle _0} + \left\langle H \right\rangle _\Phi ^{\left( 1 \right)}, 
\label{wc-60}
\end{equation}
where ${\left\langle H \right\rangle _0} = \left\langle 0 \right|H\left| 0 \right\rangle$, whereas the $\left\langle H \right\rangle _0^{\left( 1 \right)}$ term is given by 
\begin{equation}
\left\langle H \right\rangle _\Phi ^{\left( 1 \right)} =  - \frac{1}{2}\left\langle \Phi  \right|\int {{d^3}x} {\Pi ^i}\frac{{\left( {{\nabla ^2} - m_Z^2} \right)}}{{\left( {\alpha {\nabla ^2} - m_Z^2} \right)}}{\Pi _i}\left| \Phi  \right\rangle,  \label{wc-65}
\end{equation}

Now making use of equation (\ref{wc-55}) and following our earlier procedure, we find that the potential for two opposite charges, located at ${\bf y}$ and ${\bf y}^{\prime}$, takes the form 
\begin{equation}
V =  - \frac{{{q^2}}}{{4\pi }}\frac{1}{\alpha }\frac{{{e^{ - ML}}}}{L} + \frac{{{q^2}m_Z^2}}{{8\pi \alpha }}\ln \left( {1 + \frac{{{\Lambda ^2}}}{{{M^2}}}} \right)L. \label{wc-70}
\end{equation}
where ${M^2} = \frac{{m_Z^2}}{{\left( {1 + 2{\chi ^2}{{\bf v}^2}} \right)}}$, $|{\bf y} - {{\bf y}^ {\prime} }| \equiv L$ and $\Lambda$ is a cutoff. This result explicitly shows the effect of including the interaction term (\ref{wc-01}) in the model under consideration. In fact, we see that the static potential profile displays the conventional screening part, encoded in the Yukawa potential, and the linear confining potential. Before to proceed further, we would like to illustrate how to give a meaning to the cutoff $\Lambda$. To this end one recalls that that our effective model for the electromagnetic field is an effective description that arises after integrating over the $Z_{\mu}$-field, whose excitation is massive. Accordingly, ${{l_Z}}=\frac{1}{{{m_Z}}}$, the Compton wavelength of this excitation defines a correlation distance. Thus, physics at distances of the order or lower than  
${1 \mathord{\left/{\vphantom {1 {{m_Z}}}} \right.\kern-\nulldelimiterspace} {{m_Z}}}$
must take into account a microscopic description of the $Z$-fields. To be more precise, if we work with energies of the order or higher than $m_{Z}$, our effective description with the integrated effects of $Z$ is no longer sensible. In view of the foregoing remark, we can identify $\Lambda$ with $m_{Z}$. Thus, finally we end up with the following static potential profile:
\begin{equation}
V(L) =  - \frac{{{q^2}}}{{4\pi }}\frac{1}{\alpha }\frac{{{e^{ - ML}}}}{L} + \frac{{{q^2}m_Z^2}}{{8\pi \alpha }}\ln \left( {1 + \frac{{{m_{Z} ^2}}}{{{M^2}}}} \right)L. \label{wc-75}
\end{equation}

From equation (\ref{wc-75}) it follows that the corresponding interparticle force reads
\begin{equation} 
 F(L) =  - \frac{{{q^2}}}{{8\pi \alpha }}\left\{ {\frac{2}{{{L^2}}}\left( {1 + \frac{L}{{\sqrt \alpha  {l_z}}}} \right){e^{ - \frac{L}{{\sqrt \alpha  {l_z}}}}} + \frac{{\ln \left( {1 + \alpha } \right)}}{{l_z^2}}} \right\}. \nonumber\\
 \label{wc-80} 
\end{equation}

With the foregoing information provided by the interparticle one-photon-exchange potential and the corresponding force, we can proceed to present and discuss some estimates that confirm the consistency of our result.
We first notice the appearance of two length scales in the potential $V$, namely,  $l_M = 1/M $,  and the $Z^0$'s Compton wavelength, $l_Z = 1/m_Z \, (\sim 2 \times 10^{-18} m)$;  let us not forget the relationship $l_M = \sqrt{\alpha} \, l_z > l_z$, since $ \alpha > 1$.  By adopting the current parametrization given in the literature \cite{Gounaris:1999kf,Green:2016trm}, our effective coupling in the anomalous $Z \gamma \gamma$-vertex, $\chi$, is actually given by $\chi = - \,e h_3^\gamma / m_z^2$, where $h_3^\gamma$  can be extracted from $Z \gamma$-producting processes. According to the results in the paper of Ref. \cite{Green:2016trm}, $h_3^\gamma$ is estimated to be of order $10^{-3}$. 

Since $\alpha = 1 + 2 \chi^2 \, {\bf B}^2$ (we are here taking ${\bf v} = {\bf B}$, the external magnetic field), even for the strongest magnetic field that may induce instabilities in the electroweak vacuum, which is estimated to be of the order of $10^{19} \, T$, the quantity $\alpha =1 + 2 \times 10^{-8}$. Higher values of $\alpha$ are possible only for magnetic fields that produce high instabilities in the Higgs vacuum and therefore the electroweak scenario is no longer consistent. It is of interest also notice that, according to Vachaspati \cite{Vachaspati1,Vachaspati2}, magnetic fields of the order of $10^{19}\,T$ were generated in the Early Universe, during the electroweak phase transition (EWPT), when,  we can say, the electromagnetic field was born, i.e.,  appeared as the combination of the neutral weak-isospin and the weak-hypercharge gauge potentials. Though magnetic fields that strong are much above the Schwinger critical value $(B_{cr} \sim 4.41 \times 10^{9}\, T)$, let us consider the scenario of the EWPT and make below some considerations about the confining component of the interparticle potential we have derived. 
According to the results reported by Ambjorn and Olesen in the papers of \cite{Olesen1,Olesen2}, the EWPT is a slow process. They actually conclude that the Higgs field may have reached the present value of its vacuum expectation value only at around the QCD phase transition (QCDPT). This, in turn, means that the $Z$- and the $W$-gauge bosons reached their masses of $91.2$ GeV and $80.4$ GeV, respectively, only at the QCDPT, and not in the immediate aftermath of the EWPT. At the stage of the QCDPT, however, the EWPT magnetic field has lowered down to $10^{16}$ T \cite{Delia,Tevzadze,Chernodub1,Chernodub2}, so that our $\alpha$-parameter is given by  $\alpha = 1 + 2 \times10^{-14}$, for the sake of our estimates. So, from now on, we shall restrict our considerations to the vicinity of the QCDPT and $m_{Z} = 91.2$ GeV.

\section{Discussion}

By inspecting the expressions for both the interparticle potential and force, and assuming a magnetic field of $10^{19}$ T, we get that the Yukawa-type force dominates over the confining component up to distances $L = 1.33 \, l_Z  \sim  2.66 \times 10^{-18} \, m$. For $L > 1.33 \, l_Z$, the (constant) confining force is the dominant one. We have estimated the total (attractive) force for $L = 1.33 \, l_Z$; the result is  $F = 2.58 \times 10^6 N$. For distances $L$ of the order $10 \, l_Z$  and larger, the Yukawa-like force becomes negligeable and the constant confining force converges to $1.28 \times 10^6 N$. Just for the sake of comparison, let us recall that, by considering the purely classical Coulombian scenario for a quark-antiquark pair, the attractive force would be of $\mathcal{O}\left( {{{10}^7}N} \right)$. The force we have attained in our effective approach is one order of magnitude weaker because the interaction under consideration is shielded by the effect of the mass of the $Z$-particle.

This result is particularly interesting if we consider the interaction between a quark – antiquark pair. The distance $2.66 \times 10^{-18} \, m$ is around three orders of magnitude lower than the typical nucleous radius and our estimate indicates that the anomalous parity-preserving $Z \gamma \gamma$-vertex we are investigating yields an effective electroweak confining force that contributes to the (much stronger) colour-confining force. But, what we wish to highlight is that this anomalous vertex enhances confinement.

It should be emphasized again that after integrating over the $Z_{\mu}$-field and deriving the particle-antiparticle interaction potential, it is mandatory to discuss the physical scenario where it may be applied. Let us keep in mind that the $Z_{\mu}$-mass sets up the energy scale and its corresponding Compton wavelength, $m_{Z}^{-1}$, naturally introduces a length scale. We claim that, in this special case of the $Z_{\mu}$-field integrated over, our particle-antiparticle potential cannot be applied to the charged leptons (electron, muon and tau) and their corresponding antiparticles, and neither to the lighter quarks, like  $u$, $d$, $s$, $c$, $b$. Their respective Compton wavelengths are (all expressed in fm): $1.3 \times 10^{2}$, $3.9 \times 10$, $2.1$, $1.5 \times 10^{-1}$ and $4.1 \times 10^{-2}$. For the $Z$-boson, the Compton wavelength is $2.2 \times 10^{-3}$ fm. On the other hand, the $t$-quark is the only quarks whose Compton wavelength is smaller if compared with the $Z$'s : $1.1 \times 10^{-3}$ fm. With these data in our hands, it is clear that the top is the only quark that does not probe the tiny microscopic effects of the $Z$ gauge boson, and so it is a reliable approximation to integrate over the effects of the $Z^{\mu}$-field to derive the quark-antiquark interparticle electrostatic potential from our effective photonic Lagrangean. 
 With this picture in mind, for the sake of the $t$-quark, the perturbative vacuum is no longer trivial; it rather behaves as an effective vacuum which incorporates the $Z$-boson effects by integrating over the $Z^{\mu}$-field coupled to the photon field through the neutral $3$-vertex coupling. As for the charged leptons and the lighter quarks, their Compton wavelengths are long enough and they spread in such a way that the microscopic effects of the $Z$-particle cannot be neglected, so that the integration over $Z^{\mu}$ is not a sensible approximation. As a consequence, the scenario of a confining electrostatic potential as the one we have worked out here does not apply for the charged leptons and the $u$-, $d$-, $s$-, $c$- and $b$- quarks. But, if we wish to consider this framework for the top, we face a problem with the very short top-quark's meanlife time, $5 \times 10^{-25}$ s, one order of magnitude below the hadronisation time scale, which is roughly $10^{-24}$ s. This is why the top does not hadronise. However, it is relevant for this scenario to notice that our interparticle potential carries intrinsic length and time scales, respectively the $Z$'s Compton wavelength given above and meanlife time, $2.6 \times 10^{-25}$ s. This means that, before the top decays into $b\,+ \,W^{+}$, it experiences the top-antitop electrostatic potential from which we have estimated a constant confining force presented previously. We should however clarify that it is not the confining component of the electrostatic potential the responsible for the top decay; the latter is due to the weak force. We are simply arguing that, before the top weakly decays, it feels the effect of the confining force calculated above. The time scale associated to the interparticle potential derived above ensures the top itself feels the electrostatic confining force before it decays.

Finally, we also draw attention to the fact that in the ${v^{i0}} = 0$ and ${v^{ij}} \ne 0$ case, equation (\ref{wc-20}) can be brought to the form 
\begin{eqnarray}
{\cal L} &=&  - \frac{1}{4}{f_{\mu \nu }}\left[ {1 - {\chi ^2}{v^{ij}}{v_{ij}}\frac{{{\nabla ^2}}}{{\left( {{\nabla ^2} - m_Z^2} \right)}}} \right]{f^{\mu \nu }} \nonumber\\
&+& \frac{{{\chi ^2}}}{2}{v^{kj}}\left( {{\partial ^i}{f_{ij}}} \right)\frac{1}{{\left( {{\nabla ^2} - m_Z^2} \right)}}{v_{nk}}\left( {{\partial _m}{f^{mn}}} \right).\label{wc-200}
\end{eqnarray}
Proceeding in the same way as was done for the magnetic case, we find that the static potential turns out to be equation (\ref{wc-75}), in this case $\alpha  = 1 - \frac{{{\chi ^2}}}{2}\left\langle {{{\bf E}^2}} \right\rangle$ and ${M^2} = \frac{{m_Z^2}}{{\left( {1 - \frac{{{\chi ^2}}}{2}\left\langle {{{\bf E}^2}} \right\rangle } \right)}}$.

This result shows that, in the situation of an external electric field, the previous results for a magnetic field hold through, the difference, however, lies in the expression for $\alpha$, which is now no longer bigger than $1$. Contrary, now, $\alpha < 1$. It might happen that $\alpha = 0$ if  $E^{2} = 2 \chi^{-2}$. By taking the value for $\chi$ we have considered in the magnetic case, $E$ turns out to be of order $10^{8} \, GeV^{2}$, in international units $E \sim 10^{31}\,Vm^{-1}$, which is a huge electric field, $13$ orders of magnitude above the critical Schwinger field, namely, $10^{18}\, Vm^{-1}$. Actually, electric fields such that $eE \sim m_{e}^{2}$  have enough energy to decay and produce the formation of an $e^{-}e^{+}$-pair. But, this would not be compatible with our physical scenario of an interparticle potential. The creation of $e^{-}e^{+}$-pairs spoils our potential approximation. Electric fields such that $eE \sim m_{e}^{2}$  are just of the order of the critical Schwinger electric field. So, we are bound to consider external fields below $10^{18}\,Vm^{-1}$. In such a case, $\alpha < 1 + {\mathcal O}(10^{-28})$. Even at the scenario of the QCDPT, with electric fields of the order of $10^ {21} \,Vm^{-1}$, there is no risk of reaching the $\alpha=0$ - singularity. Anyway, in our considerations below, we shall be inspecting the interparticle potential for lower fields, as we shall specify. Actually, electric fields such that $eE \sim m_{q}^{2}$ ($m_{q}$ is the quark mass) have enough energy to decay and to produce the formation of a quark-antiquark pair. But, this would not be compatible with our physical scenario of an interparticle potential. (The creation of particle-antiparticle pairs spoils our potential approximation.) Electric fields such that $eE \sim m_{t}^{2}$  ($m_{t}$ is the top quark mass) are of the order of $10^{28} \, Vm^{-1}$. However, since our scenario is the one of QCDPT, the electric fields involved are seven orders of magnitude below $10^{28}\, Vm^{-1}$, so they are not strong enough to produce top-antitop pairs, which would invalidate our potential approximation. By then considering $E \sim 10^{21}\, Vm^{-1}$ and keeping the value $10^{-8}\, GeV^{-2}$ for $\chi$, the $\alpha$-parameter is such that $\alpha < 1 + {\mathcal O}(10^{-22})$. So, for the sake of our estimations, we can still undertake that, as in the case of external magnetic fields, the presence of external electric fields typical of the QCDPT, which are not able to create top-anti-top pairs, the anomalous tri-vertex we are investigating keeps on enhancing confinement.

 \section{Final remarks}

In summary, using the gauge-invariant but path-dependent formalism, we have computed the interaction energy when an anomalous triple gauge boson couplings is taken into account in the electroweak sector of the Standard Model, and  the $SU(2)_{L}\otimes U(1)_{Y}$ symmetry is spontaneously broken by the Higgs mechanism to $U(1)_{em}$. Once again, a correct identification of physical degrees of freedom has been crucial for understanding the physics hidden in gauge theories. Interestingly, it was shown that the interaction energy is the sum of a Yukawa and a linear potential, leading to the confinement of static probe charges.\\

\section{acknowledgements}

The authors are grateful to Prof. T. Vachaspati for a fruitful communication and references.
One of us (P. G.) was partially supported by Fondecyt (Chile) grant 1180178 and by ANID PIA / APOYO AFB180002. L. P. R. O. is grateful to the PCI-DB funds (CNPq / MCTI).


\begin{thebibliography}{99}

\bibitem{Atlas} ATLAS Collaboration, G. Aad, et al., Phys. Lett. B {\bf 716}, 1 (2012).

\bibitem{CMS} CMS Collaboration, S. Chatrchyan, et al., Phys. Lett. B {\bf 716}, 30 (2012).

\bibitem{Englert} F. Englert and R. Brout, Phys. Rev. Lett. {\bf 13}, 321 (1964).
 
\bibitem{Higgs1} P. W. Higgs, Phys. Lett. {\bf 12}, 132 (1964). 

\bibitem{Higgs2} P. W. Higgs, Phys. Rev. Lett. {\bf 13}, 508 (1964).

\bibitem{Rahaman:2018ujg} R.~Rahaman and R.~K.~Singh, Nucl. Phys. B \textbf{948}, 114754 (2019).

\bibitem{Senol:2018cks} A.~Senol, H.~Denizli, A.~Yilmaz, I.~Turk Cakir, K.~Y.~Oyulmaz, O.~Karadeniz and O.~Cakir, Nucl. Phys. B \textbf{935}, 365-376 (2018).

\bibitem{Senol:2019qyl} A.~Senol, H.~Denizli, A.~Yilmaz, I.~Turk Cakir and O.~Cakir, Phys. Lett. B \textbf{802}, 135255 (2020).

\bibitem{Larios:2000ni} F.~Larios, M.~A.~Perez, G.~Tavares-Velasco and J.~J.~Toscano, Phys. Rev. D \textbf{63}, 113014 (2001).

\bibitem{Degrande:2013kka} C.~Degrande, JHEP \textbf{02}, 101 (2014).

\bibitem{Hagiwara:1986vm} K.~Hagiwara, R.~D.~Peccei, D.~Zeppenfeld and K.~Hikasa,
Nucl. Phys. B \textbf{282}, 253-307 (1987).

\bibitem{Gounaris:1999kf} G.~J.~Gounaris, J.~Layssac and F.~M.~Renard, Phys. Rev. D \textbf{61}, 073013 (2000).

\bibitem{Green:2016trm} D.~R.~Green, P.~Meade and M.~A.~Pleier, Rev. Mod. Phys. \textbf{89}, no.3, 035008 (2017).

\bibitem{Gaete:2017vxk} P.~Gaete and J.~A.~Helay\"{e}l-Neto, EPL \textbf{120}, no.1, 11001 (2017).

\bibitem{Gaete:2017cpc} P.~Gaete and J.~A.~Helay\"{e}l-Neto, EPL \textbf{119}, no.5, 51001 (2017).

\bibitem{Gaete:1997eg} P.~Gaete,  Z.\ Phys.\ C {\bf 76}, 355 (1997).

\bibitem{Vachaspati1} T. Vachaspati, Phys. Lett. {\bf B265} (1991) 258.

\bibitem{Vachaspati2} T. Vachaspati, Phil. Trans. R. Soc. {\bf A366} (2008) 2915.

\bibitem{Olesen1} J. Ambjorn and P. Olesen, Int. J. Mod. Phys. {\bf A5} (1990) 4525. 

\bibitem{Olesen2} J. Ambjorn and P. Olesen, Nucl. Phys. {\bf B330} (1990) 193. 

\bibitem{Delia} M. D'Elia, S. Mukherjee and F. Sanfilippo, Phys. Rev. {\bf D82} (2010) 051501.

\bibitem{Tevzadze} A. G. Tevzadze, L. Kisslinger, A. Brandenburg and T. Kahniashvili, Astrophys. J. {\bf 759} (2012) 54.

\bibitem{Chernodub1}  M. N. Chernodub, Van J. Doorsselaere and H. Verschelde, Phys. Rev. {\bf D88} (2013) 065006.

\bibitem{Chernodub2} M. N. Chernodub, Int. J. Mod. Phys. {\bf D23}, (2014) 1430009.
\end{thebibliography}
\end{document}